\documentclass[aps,twocolumn,showkeys,groupedaddress,prb]{revtex4}
\usepackage{graphicx}
\usepackage{amsmath}
\usepackage{amssymb}
\begin{document}

\title{Range of ion specific effects in the hydration of ions}
\author{Safir Merchant}
\author{Dilip Asthagiri}\thanks{Corresponding author: Fax: +1-410-516-5510; Email: dilipa@jhu.edu}
\affiliation{Department of Chemical and Biomolecular Engineering and The Institute of NanoBioTechnology,  Johns Hopkins University, Baltimore, MD 21218}

\date{\today}
\begin{abstract}
Within the quasichemical approach, the hydration free energy of an ion is decomposed into a chemical term accounting for 
ion specific ion-water interactions within the coordination sphere and nonspecific contributions accounting for packing (excluded volume) and long range interactions. The change in the chemical term with a change in the radius of the coordination sphere is the compressive force exerted by the bulk solvent medium on the surface of the coordination sphere. For the Na$^+$, K$^+$, F$^-$, and Cl$^-$ ions considered here this compressive force becomes equal for similarly charged ions for coordination radii of about 3.9~{\AA}, not much 
larger than a water molecule. These results show that ion specific effects are short ranged and arise primarily due to differences in the local ion-water interactions.
\end{abstract}  
\keywords{Hofmeister effect, potential distribution theorem, Monte Carlo, simulations}

\maketitle

Numerous aqueous phase processes including, for example, the stability of biological macromolecules and 
colloidal suspensions,  are strongly influenced by the chemical type of the inorganic ions dissolved in the aqueous phase  \cite{kunz:cocis04}. However, despite their ubiquity, a molecular level understanding of ion specific effects remains elusive\cite{kunz:cocis04,zhang:cocb06}. Ideas such as distortion of hydrogen bond network of the bulk solvent\cite{marcus:chemrev09}, balance between ion-water and water-water interactions\cite{collins:meth04}, and influence of ion-specific dispersion forces \cite{ninham:langmuir97} have been proposed to rationalize these effects.  

Understanding how the ion affects the solvent matrix remains of first interest in understanding ion specific effects. 
Recent spectroscopic measurements on salt  solutions\cite{omta:sci03, kropman:cpl03, bakker:jpcond05}  suggest that 
the ion only affects hydrogen bonds of water molecules within its first hydration shell. An earlier theory and simulation study 
by us \cite{merchant:jcp09} also showed that only a small subset of water molecules in the first hydration shell of the ion are sensitive
to the type of the ion.  These results taken together suggest that ion specific ion-solvent interactions are limited only to the ion's local neighborhood.  In this Communication, we address the question of how far in the liquid the chemical type of the ion is felt and if the 
point at which ion-specificity is lost is also the point where continuum models of hydration begin to take hold. 

To address these questions, we first define a coordination sphere of radius $r$ around the ion to separate the ion interaction with
water molecules within the coordination shell from the longer-range interaction of the ion with the solvent outside the coordination 
sphere. With such a spatial partitioning, the hydration free energy of the ion, $\mu^{\rm ex}$, can be written as\cite{lrp:ES99, lrp:apc02, lrp:book, lrp:cpms}\begin{eqnarray}
 \mu^{\rm ex} &=&  k{\rm_B}T\ln x_0(r)  - k{\rm_B}T\ln p_0(r) +  \mu^{ex}_{\rm outer}(r) \, .
\label{eq:qc}
\end{eqnarray} 
where $k_{\rm B}$ is  Boltzmann's constant and $T$ is the temperature,  $k{\rm_B}T\ln x_0(r)$ is the chemical contribution to the
hydration free energy, $-k{\rm_B}T\ln p_0(r)$ is the packing contribution, and $ \mu^{ex}_{\rm outer}(r)$ is the contribution to the hydration free energy due to ion interaction with the solvent outside the coordination shell.  $x_0(r)$ is the probability of observing no water molecules within the coordination sphere of radius $r$ around the ion; thus $k{\rm_B}T\ln x_0(r)$ is the free energy gained by allowing water molecules to flood the empty coordination sphere, justifying the identification of this term with local, chemical interactions. $p_0(r)$ is the probability of observing an empty coordination sphere, but in the absence of the ion; $-k{\rm_B}T\ln p_0(r)$ is the free energy required to create a cavity of radius $r$ in the bulk solvent. $\mu^{ex}_{\rm outer}(r)$ is the free energy gained in placing the ion in an empty cavity; it accounts for the interaction of the ion with the bulk medium outside the coordination sphere.  For a sufficiently large $r$, we expect $\mu^{\rm ex}_{\rm outer}(r)$ to be composed of a large number of small non-specific contributions \cite{asthagiri:jacscf4,shah:jcp07}. Since Eq.~\ref{eq:qc} is a tautology, 
the point where $\mu^{\rm ex}_{\rm outer}(r)$ becomes insensitive to ion type is also the range to which $k_{\rm B}T \ln x_0(r)$  captures all the ion-specific effects. 

To identify the range of ion specific effects, we studied ion-water systems under NVT conditions using Metropolis Monte Carlo simulations\cite{metropolis:jcp53, allen}. Water was modeled using the SPC/E model\cite{berendsen:jpc87} and parameters for Na$^+$, K$^+$, F$^-$, Cl$^-$ ions were obtained from an earlier study\cite{hummer:jpc96}. The chemical term was estimated by growing the
coordination sphere in steps of 0.05~{\AA}. At any $r$, the probability of observing zero (0) water molecules in a coordination sphere of radius $r+0.05$~{\AA} was obtained. From this data, the entire $x_0(r)$ versus $r$ profile was reconstructed. A similar strategy was followed for $p_0(r)$ versus $r$, but with a step size of 0.1~{\AA}. At each $r$ value, $3\times10^5$ sweeps of equilibration were followed by $3\times10^5$ sweeps of production. Each sweep comprised one attempted move of each particle in the system. The maximum allowed translation and rotation were adjusted during the equilibration phase to target an acceptance ratio of 0.3. In the production phase these maximum displacement values were held fixed. Configurations are saved every 10 sweeps for analysis.

\begin{figure}
\includegraphics{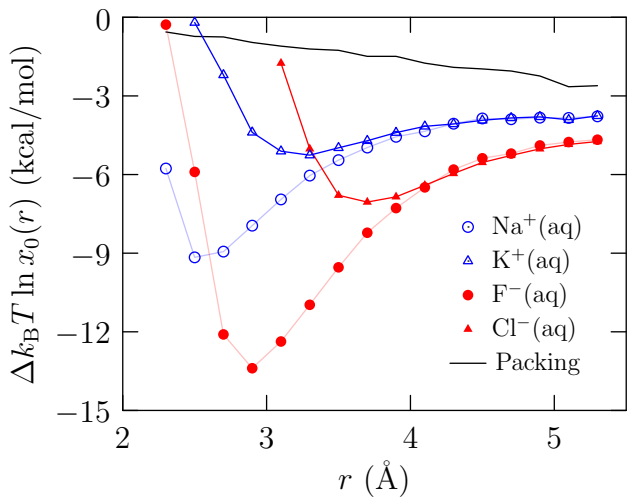}
\caption{ The change in the chemical term, $\Delta k_{\rm B}T\ln x_0$, with change in the radius, $r$, of the coordination 
shell for various ions. The dashed line is the result for a solute that does not interact with the solvent; it thus accounts for the change in the packing contribution. Observe that the change in the chemical term approaches the change in the packing term as the coordination shell is increased. For $r > 3.9$ \AA~the change in the chemical term is the same between identically charged ions.}
\label{fig:change} 
\end{figure}

From Fig.~\ref{fig:change} we observe that for $ r > 3.9$ {\AA} the change in the chemical term depends only on the charge of the ion and is independent of its chemical nature.  
Fig.~\ref{fig:change} emphasizes that, for the ions studied here, ion-specific effects are short ranged. In particular, 
for $r > 3.9$ \AA, the increase in the chemical term is due to non-specific contributions. 

To elucidate the non-specific nature of contributions for $r > 3.9$~{\AA}, we first note that 
\begin{eqnarray}
 k{\rm_B}T \frac{\partial\ln x_0(r)/p_0(r)}{\partial r}  = -  \frac{\partial\mu^{ex}_{\rm outer}(r)}{\partial r}\, .
\label{eq:qcpartial}
\end{eqnarray} 
The left hand side of the above equation is the compressive force of the bulk on the coordination sphere of the ion relative to the 
compressive force on an ion-free coordination sphere \cite{beck:jcp08}.  

For a sufficiently large $r$, we expect the interaction of the ion with the bulk medium outside the coordination sphere to be well-described by a Gaussian \cite{asthagiri:jacscf4,shah:jcp07}. In this case, for an ion of charge $q$, we have 
\begin{eqnarray}
\mu^{\rm ex}_{\rm outer} = q \langle \phi\rangle_0 - \frac{\beta}{2}\cdot q^2 \cdot \langle (\phi - \langle \phi\rangle_0)^2 \rangle_0 \, ,
\label{eq:quad}
\end{eqnarray}
where $\langle \ldots\rangle_0$ denotes averaging in the presence of an uncharged ion, $\langle \phi\rangle_0$ is the potential at the center of an uncharged ion, and $\langle (\phi - \langle \phi\rangle_0)^2 \rangle_0$ is the fluctuation in the electrostatic potential \cite{hummer:jpc96}. By defining $1/r_{\rm Born} = \beta \langle (\phi - \langle \phi\rangle_0)^2 \rangle_0 / 2$, it is readily seen that the fluctuation contribution is just the Born model of hydration.  (For clarity, in Eq.~\ref{eq:quad} we do not explicitly indicate the corrections due to electrostatic self-interaction\cite{hummer:jpc96}, as these depend solely on $q$ and not on the ion type.)

In the radius range of interest we find that $\partial \langle \phi\rangle_0/\partial r$ is about a $k_{\rm B} T/ {\rm {\AA}}$. (A similar estimate is obtained based on the data in Ref.\ \onlinecite{ashbaugh:jpcb00} for a cavity.) Neglecting this contribution, we thus expect that asymptotically
\begin{eqnarray}
 k{\rm_B}T \frac{\partial \ln x_0/p_0}{\partial r} \sim -\frac{q^2}{2r^2}(1 - \frac{1}{\epsilon}) \, ,
\label{eq:mu0}
\end{eqnarray} 
where $\epsilon$ is the dielectric constant for water. 

\begin{figure}
\includegraphics{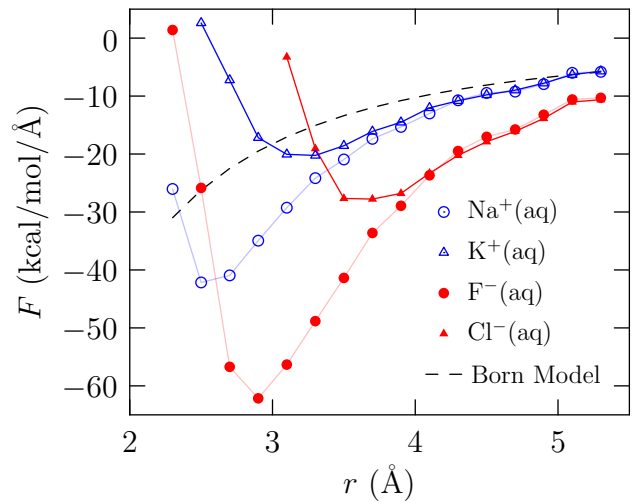}
\caption{$F = k{\rm_B}T \frac{\partial\ln x_0(r)/p_0(r)}{\partial r}$, Eq.~\ref{eq:qcpartial}, the compressive force exerted by the solvent
on the surface of the coordination shell containing the ion, $k{\rm_B}T \partial[\ln x_0(r)]/\partial r$, relative to the compressive force on an empty cavity, $k{\rm_B}T \partial[\ln p_0(r)]/\partial r$ for various coordination radii, $r$. The black dashed line is the estimate of the relative compressive force based on the Born model of hydration (Eq.~\ref{eq:mu0}).}
\label{fig:diff1} 
\end{figure}

Figure~\ref{fig:diff1} shows that for $ r > 3.9$ {\AA} the relative compressive force between similarly charged ions becomes
equal, but is different for anions and cations.  The large $r$ value is also different from the value obtained 
using the Born model, where the Born-radius is set equal to the coordination radius. 
\begin{figure}
\includegraphics{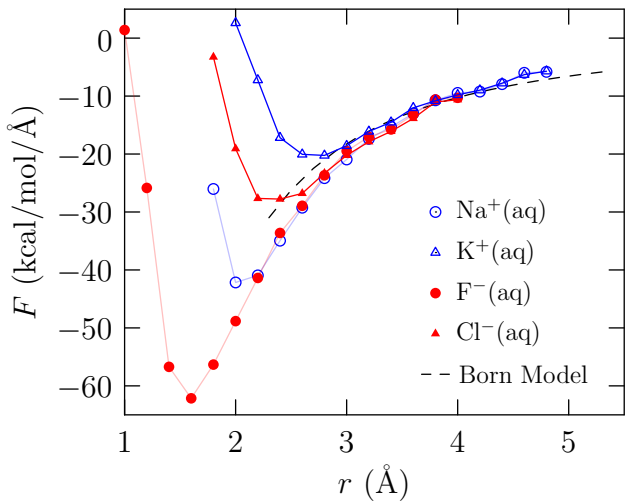}
\caption{The data in Fig.~\ref{fig:diff1} have been shifted along $r$ by $-1.3$ {\AA} for the anions and by $-0.5$ {\AA} for the cations. With this empirical shift,  the compressive force $F = k{\rm_B}T \frac{\partial\ln x_0(r)/p_0(r)}{\partial r}$, Eq.~\ref{eq:qcpartial}, 
follows the trend predicted by the Born model (Eq.~\ref{eq:mu0}).}
\label{fig:diff2} 
\end{figure}   

The discrepancy in the asymptotic limit between simulation data and that based on the Born model is not particularly surprising, 
since the coordination radius is not the same as the Born radius \cite{hummer:jpc96,ashbaugh:jpcb00}. For example, 
the position of the water molecule is denoted by the position of its oxygen atom and thus the coordination shell strictly excludes only the oxygen atoms of the water molecules but not their corresponding hydrogen atoms, which can enter the coordination sphere. Since this
effect will be more pronounced for the anions than the cations, we expect that the equivalent Born radius for anions to be smaller than that for the cations. What is interesting, however, is the observation that by empirically reducing the cavity radius by 1.3 \AA~for anions and 0.5 \AA~for cations (Fig.~\ref{fig:diff2}), the relative compressive force for both anions and cations are similar for large coordination radii. Further,  the large-$r$ curvature of the radially translated curves compares reasonably well with that based on the the Born model of hydration. (We maintain the difference of 0.8 \AA, as earlier studies on such shifts find a similar difference for anions and cations\cite{lps:jcp39, ashbaugh:jcp08}.) 

In summary, for the ions studied here, we find that a coordination radius of about 3.9~{\AA}, not much larger than the size of a water molecule, is sufficient to account for all ion specific ion-water interactions.  Our results are in accordance with experimental results\cite{omta:sci03, kropman:cpl03, bakker:jpcond05}  that suggest that the influence of the ion on hydrogen bonding between water molecules 
is minimal for water molecules beyond the first hydration shell. The short-range of ion specific ion-water interactions further
suggests that any framework for modeling ion-specific effects needs to acknowledge the molecular characteristics of local ion-water 
interactions.

% \bibliography{metals}

 \end{document}